In a recent letter, Gaidarzhy *et al.* [1] claim to have observed evidence for "quantized displacements" of a nanomechanical oscillator. We contend that the evidence, analysis, claims, and conclusions presented are contrary to expectations from fundamentals of quantum mechanics and elasticity theory, and that the method used by the authors is unsuitable in principle to observe the quantized energy states of a nanomechanical structure.

1) According to standard quantum mechanics, measurement of energy quanta in a resonator mode requires a measurement probe whose interaction Hamiltonian commutes with the oscillator Hamiltonian, i.e., one requires a quantum non-demolition (QND) measurement scheme. However, with continuous linear measurement of position or velocity, the best energy sensitivity one can in principle achieve is: $\Delta E \approx \hbar \omega \sqrt{\overline{N}}$, where $\overline{N}$ is the average number of quanta, the so-called standard quantum limit (SQL) [2].

The authors incorrectly claim to measure the absolute value of position. They actually employ continuous magnetomotive detection [3]. This is well understood as a continuous linear measurement scheme and is not a QND measurement of the energy in any approximation. In magnetomotive detection, the sample is immersed in a large magnetic field, and driven with an oscillating current through the mechanical element. The magnetic field transforms the applied oscillating currents into forces on the resonator, and transduces the resulting mechanical oscillations into measurable voltages. The authors use a room temperature semiconducting amplifier with a noise temperature of $T_N=440K$ to detect these voltages [4]. Thus, in addition to the magnetic drive, the backaction current noise of such an amplifier will drive the resonator, acting as a thermal bath far above temperature of 100mK quoted by the authors.

The authors fail to point out that the resonator is driven many orders of magnitude above the ground and first excited state during the measurement shown in Fig. 4c. Given the reported parameters (F ~ 45 pN, $k_{eff}$ = 188 N/m, and Q=150), the average number of energy quanta in the resonator during the measurement is $\overline{N} = 120,000 \gg 1$ which corresponds to an effective resonator temperature of 8800K. Even with displacement detection at the SQL it is in principle impossible to resolve the energy quanta. Given this drive amplitude, one does not expect to observe any evidence of the lowest quantized energy states of the resonator.

2) For Q~100 and $\omega = 10^{10}$ s$^{-1}$, the average lifetime of an energy quantum is ~10 nsec. Even if the authors could measure the oscillator energy with single quantum accuracy, the observed jumps due to decay would certainly not be as long as tens of minutes, which is the time scale indicated on Fig. 4c, a discrepancy of over 10 orders of magnitude from expectation.

3) The magnetomotive response of the suspected 1.48GHz mode is anomalous. Fig. #3b shows the amplitude versus magnetic field where the authors claim that it demonstrates the expected quadratic dependence. The authors fail to point out and explain that the quadratic fit is not symmetric about the origin as is expected and widely observed, and appears to fit a parabola offset by -2 Tesla. This behavior has not been observed in other groups' measurements of similar frequency resonators at low temperature.

Finally, there is no justification offered for the assertion that the motion of the central beam is amplified from femtometers to picometers in comparison to the femtometer motion of the fingers at 1.5GHz. The mode that the authors claim they are observing is a "flapping" mode of the structure, in which the fingers move coherently and out of phase with the central support. Both the simulation shown by the authors and those performed by us indicate that the central beam moves with displacements which are similar in magnitude to that of the fingers. In principle, no displacement amplification of the quantum motion is possible by a passive structure: the result of making the structure larger to contain multiple sub-elements yields smaller quantum fluctuations since the quantum of energy, $\hbar\omega$, must be distributed over a larger mass structure.

In summary, we find no evidence that the magnetomotive impedance jumps, which Gaidarzhy *et al.* observe by driving their resonator to very high amplitude, have any connection to quantum phenomena. Their assertion that these originate from the quantized excitations of a mechanical resonator are in direct conflict with basic quantum mechanics, simple elasticity theory, and the (expected) behavior observed with all other similar nanomechanical structures reported to date.

K.C. Schwab, Laboratory for Physical Sciences
M.P. Blencowe, Dartmouth College
M.L. Roukes, California Institute of Technology
A.N. Cleland, University of California, Santa Barbara
S.M. Girvin, Yale University

G.J. Milburn, University of Queensland
K.L. Ekinci, Boston University